\documentclass[twocolumn,prb,aps,showpacs]{revtex4}
\def\half{{1\over 2}}
\def\cm{c_-}
\def\vph{v_{\rm ph}}
\def\const{\mbox{const}}
\def\rate{{\cal R}}
\def\volt{V}
\def\Im{\mbox{Im}}
\def\prop{{\cal G}}
\def\be{\begin{equation}}
\def\ee{\end{equation}}
\def\ba{\begin{eqnarray}}
\def\ea{\end{eqnarray}}
\begin{document}
\title{Piezoelectric coupling, phonons, and tunneling into a quantum Hall edge}
\author{S. Khlebnikov}
\affiliation{Department of Physics, Purdue University, West Lafayette,
Indiana 47907, USA}
 
\date{December 14, 2005}

\begin{abstract}
We show that the piezoelectric coupling to three-dimensional phonons in GaAs
renormalizes the current-voltage exponent for tunneling of electrons into an
incompressible quantum Hall edge. The leading correction is always negative, in 
agreement with experiments on the $\nu = 1/3$ state and, depending on the precise 
value of the edge plasmon speed, can be as large as a few percent. We also 
discuss higher-order corrections, which determine the effect of the
piezoelectric coupling in the extreme infrared limit.

\end{abstract}
 
\maketitle
\section{Introduction}
The discrepancy between the tunneling exponent predicted by the conformal
field theory\cite{Wen} (CFT) of a quantum Hall (QH) edge and the one measured in 
cleaved-edge-overgrowth (CEO) experiments\cite{Chang&al,Grayson&al}, for a review
see Ref. \onlinecite{Chang},
remains a puzzle. For the principal filling fraction $\nu = 1/3$,
the discrepancy is not that large but is still believed to require
an explanation (for recent discussions of the issue, see Refs. 
\onlinecite{Boyarsky&al,Wan&al2005}).
A natural way to resolve the discrepancy would be to identify additional gapless
modes, not present in the original chiral CFT. Such modes
appear, for example, in scenarios based on 
the ``edge reconstruction''.\cite{Chamon&Wen,Wan&al2002} 
Another group of explanations involves the role of the Coulomb interaction in the
presence of a ``hard'' edge;\cite{Tsiper&Goldman,Mandal&Jain} profiles of the 
electron density obtained numerically in this case are rather similar to those
resulting from edge reconstruction. Finally, although we do not address that
case in the present paper, we note that there seems to be a discrepancy between
theory and experiment also for compressible QH states; for more detail, see
Refs. \onlinecite{Chang,Levitov&al}.

The effect of a one-dimensional (1D) ``phononlike'' mode on scattering between two
QH edges was considered in Ref. \onlinecite{RH}, and it was shown that a derivative 
coupling of the charge density
to such a mode does renormalize the resistance. 
For tunneling from a bulk metal into a single edge, as in the CEO experiments,
this coupling will similarly
renormalize the tunneling exponent. The requisite 1D phononlike mode, propagating
only along the edge, can be a result of the edge reconstruction or other mechanisms
enumerated in Ref. \onlinecite{RH}, all of which have to do with the near-edge
properties of the electron density.

For the purpose of explaining the experimental data, however, one does not have
to consider the electronic system alone. In particular, in the present paper
we note a hitherto apparently unidentified effect of the ordinary three-dimensional
(3D) acoustic phonons in GaAs. The derivative coupling of the type considered in
Ref. \onlinecite{RH} is irrelevant in this case, in the sense that it does not lead
to a logarithmic correction to the propagator of the edge plasmon.
(The replacement
of the 3D phonon with a 1D phonon in this problem
in Ref. \onlinecite{Heinonen&Eggert} looks to us completely arbitrary.)
As we show here, however, there is a logarithmic effect (and consequently a 
renormalization of the tunneling exponent) due to the piezoelectric coupling.

We limit ourselves to the
principal filling fractions $\nu = 1/(2p+1)$ and assume that the CFT contains 
a single chiral boson---the left-moving edge plasmon mode.\cite{Wen}
However, the phonon-induced correlation to which we attribute our results
may also play a role in a much broader class of problems involving 1D conductors.

The piezoelectric electron-phonon interaction is described by the Hamiltonian
\be
H_{\rm piezo} = h \partial_x \theta_L \phi \delta(y) \delta(z) \; .
\label{Hp}
\ee
Here $\theta_L$ is the chiral field of the edge plasmon, $\frac{1}{2\pi} \partial_x 
\theta_L$ is the representation of the electron density in the edge CFT,
$\phi$ is the phonon field, and $h$ is the coupling constant. The delta-functions
restrict the coupling to one dimension, and indeed $\theta_L$ is a 1D field;
however, $\phi$ is fully three-dimensional. 

The piezoelectric coupling is known to be highly anisotropic, but here we follow 
a common practice and use a direction average. 
The main result---a renormalization of the tunneling exponent---does not depend on 
this replacement. In fact, one
could get rid of the averaging directly in the solution, Eq. (\ref{F}) below, 
by bringing $h^2$ under the integral over $k_\perp$ (the transverse wavenumber of
the phonon field)
and supplying it with an angular dependence. This would not affect the structure of
the infrared-sensitive terms.

We keep 
in mind that there are three polarizations of acoustic phonons (two transverse
and one longitudinal), but in our calculation their contributions simply add up,
so we will think of $\phi$ as representing just one of them.
Interactions of the form (\ref{Hp}) have been considered in a variety of problems
concerning GaAs structures (for recent work, see Ref. \onlinecite{Seelig&al}) 
but, to our knowledge,
not in connection with tunneling into a QH edge. 

It is straightforward to count the powers 
of momenta and see that the Hamiltonian (\ref{Hp})
allows for a logarithmic correction to the propagator of $\theta_L$. We
can anticipate the nature of the effect by noting that the Hamiltonian
gives rise to a non-local time-dependent ``potential''
between electrons, which to the leading order in $h^2$ has the form
\be
U(x,\tau) = - \frac{h^2 }{\vph(x^2 + \vph^2 \tau^2)} \; .
\label{U}
\ee
Here $\vph$ is the phonon speed, and $\tau$ the Euclidean time. 
Since this ``potential'' is attractive, we expect that it will tend to confine
the electron cloud near the tunneling site, thus leading
to a {\em decrease} in the tunneling exponent $\alpha$.

What remains, then, is to compute, using the interaction Hamiltonian (\ref{Hp}), 
the coefficient of the logarithm, to see if
the effect can be large enough to be experimentally accessible. 
In Sect. \ref{sect:corr}, we will see that it can.
The precise answer depends on the ratio
$\cm / \vph$ of the speed of the chiral mode to that of the phonon. For 
$\cm$ in the range 10$^5$--10$^6$ cm/s, the leading (in $h^2$) correction to 
the tunneling exponent
ranges from an accessible $\Delta \alpha = -0.09$ to an apparently
unobservable $\Delta \alpha = -0.003$. The correction is always {\em negative}, 
in agreement with the experiments\cite{Chang&al,Grayson&al} on the $\nu=1/3$ state.
Higher-order corrections and the $\nu=1$ state are discussed in 
Sect. \ref{sect:horder}.
Sect. \ref{sect:concl} is a summary of the results.

\section{The model and the path integral}
We start with the Euclidean action that consists of three parts: the individual
actions of the 1D chiral boson $\theta_L$ and the 3D phonon $\phi$, and the
interaction corresponding to the Hamiltonian (\ref{Hp}):
\be
S_E = S_\theta + S_\phi + \int d\tau d^3 x H_{\rm piezo} \; .
\label{SE}
\ee
A term that describes tunneling of a single electron between a Fermi-liquid and 
the QH edge will be added later. In Eq. (\ref{SE}),
\ba
S_\theta & = & \kappa \int d\tau dx [-i \partial_\tau \theta_L \partial_x \theta_L
+ \cm (\partial_x \theta_L)^2 ] \; , \label{Stheta} \\
S_\phi & = &  \int d\tau d^3x \left\{ \half (\partial_\tau \phi)^2 
+ \half \vph^2 (\partial_x \phi)^2 \right\} \; ,
\ea
$\kappa = 1 / 4\pi \nu$, and $\tau = i t$ is the Euclidean time. We denote
1D integrations by $dx$, and 3D ones by $d^3 x$. Our normalization of 
the field $\theta_L$ is such that the expansion in terms of the canonically 
normalized creation and annihilation operators has the form
\be
\theta_L = \sum_{k<0} \frac{1}{\sqrt{2\kappa |k| L}}
(b_k e^{-i\omega_- t + ikx} + b_k^\dagger  e^{i\omega_- t - ikx}) + \mbox{z.m.} \; ,
\label{theta_L}
\ee
where $L$ is the length of the edge, and $\mbox{z.m.}$ stands for the zero 
modes.\cite{Wen} In this normalization, the fermion creation operator in the
chiral CFT is
\be
\psi^\dagger = :\exp(\frac{i}{\nu} \theta_L): \; .
\label{psi}
\ee

The easiest way to obtain a path integral for the chiral field $\theta_L$ is
to separate it into the even and odd (with respect to $x$) components:
$\theta_L = \theta_e + \theta_o$. If we consider $\theta_e$ as the canonical 
coordinate, then it follows from (\ref{Stheta}) that 
\be
p = 2\kappa \partial_x \theta_o
\label{p}
\ee
is the corresponding canonical momentum. The path-integral measure can 
be written as ${\cal D} \theta_e {\cal D} p$ or, since 
the Jacobian of transformation from $p$ to $\theta_o$ does not 
depend on the field, equivalently
as ${\cal D} \theta_e {\cal D} \theta_o = {\cal D} \theta_L$.
Thus, the vacuum-to-vacuum amplitude in the presence of a current
$J(x,\tau)$ can be written as
\be
Z[J] = \int {\cal D} \theta_L e^{-S_E + \int d\tau dx J \theta_L} \; .
\label{Z}
\ee

We assume that tunneling occurs at isolated sites---impurities, or tunneling
centers, and in what follows consider just one such site, located at $x=0$.
The tunneling term $S_t$ is then
\be
S_t = \const \int d\tau [ c(0,\tau) \psi^\dagger(0,\tau) + \mbox{H.c.} ] \; ,
\label{St}
\ee
where the operator $c(0,\tau)$ destroys a spin-polarized Fermi-liquid electron 
at point $x=0$. Treating $S_t$ perturbatively, we can obtain the tunneling rate
via a version of the optical theorem, from the two-point correlator
of $\psi$. The requisite correlator is given by Eq. (\ref{Z}) with a special
choice of the current:
\be
J(x, \tau) = \frac{i}{\nu} [ \delta(x-x_1) \delta(\tau-\tau_1) -
\delta(x-x_2) \delta(\tau-\tau_2)] \; ,
\label{J}
\ee
corresponding to insertion of a fermion  at $x=x_1$, $\tau=\tau_1$ and removal 
of a fermion at $x=x_2$, $\tau=\tau_2$.
For the present purposes, we only need to consider $x_1 = x_2 = 0$, but the 
structure of the correlator is elucidated by taking general $x_1$ and $x_2$,
so we compute it for that more general case.

Note that the correlator just defined corresponds to the bosonic version of
time ordering,
\be
\prop(x, \tau) = 
\langle 0 | T_B [ \psi(x_2, \tau_2) \psi^\dagger(x_1, \tau_1) ] | 0\rangle \; ,
\label{cor}
\ee
where $x = x_2-x_1, \tau = \tau_2 - \tau_1$, and
\be
T_B [ A(\tau_2) B(\tau_1) ] = \Theta(\tau) A(\tau_2) B(\tau_1) +
\Theta(-\tau) B(\tau_1) A(\tau_2) \; ,
\ee
even though the exponentials (\ref{psi}) are in fact fermions. This is 
inconsequential since the correlator (\ref{cor}) does not occur as an internal
line in any Feynman diagram.

We consider the case of zero temperature, $T=0$. 
The optical theorem gives the tunneling rate per a unit interval of the
biasing energy $E>0$ in terms of the analytical continuation of $\prop(x, \tau)$
to real time. For tunneling into the edge, the rate is
\be
\frac{d\rate}{dE} \propto N(-E) \Im [i \int_{0}^{\infty} 
dt  e^{i E t} \prop(0, i t + \delta) ] \; ,
\label{rateE}
\ee
where $N(-E)$ is the density of states in the Fermi-liquid.
As indicated by the infinitesimal $\delta > 0$, in Eq. (\ref{rateE}) we need
the values of the correlator just below the real-$t$ axis. These can be
obtained by analytically continuing the Euclidean $\prop(x, \tau)$ from $\tau > 0$.

Because the theory (\ref{SE}) is Gaussian, the path integral (\ref{Z}) can
be computed exactly:
\be
Z[J] = \exp (\half J G J) \; ,
\label{JGJ}
\ee
where $G$ is the full Green function of the chiral boson $\theta_L$
[convolution integrals are implied in the exponent of Eq. (\ref{JGJ})].
In the absence of phonons, $G$ can be replaced by the free Green function
\be
G_0(x, \tau) = -i \nu \int \frac{d\Omega dk_x}{2\pi}
\frac{e^{-i \Omega \tau + i k_x x}}{k_x(\Omega - i \cm k_x)} \; ,
\label{G0}
\ee
which to logarithmic accuracy equals
\be
G_0(x,\tau) = \left\{ \begin{tabular}{lr} 
$\nu \ln \frac{L}{\cm\tau + ix} \; ,$  & $\tau > 0 \; ,$ \vspace{0.1in} \\
$\nu \ln \frac{L}{-\cm\tau - ix} \; ,$  & $\tau < 0 \; .$
\end{tabular} \right. 
\ee
$L$ is an infrared cutoff, any dependence on which will disappear when we use
$G_0$ (in place of $G$) in Eq. (\ref{JGJ}).

At small $E$, where $N(-E)$ tends to a constant, Eq. (\ref{rateE}) then becomes
\be
\frac{d\rate}{dE} 
\propto \Im [i\int_{0}^{\infty}  dt e^{ i E t} (it + \delta)^{-\alpha}] 
= \frac{\pi}{\Gamma(\alpha)} E^{\alpha-1} \; ,
\label{rateE2}
\ee
where $\alpha = 4\pi \kappa= 1/\nu$. The rate at a fixed biasing voltage 
$\volt$ is
\be
\rate = \int_0^{e\volt} \frac{d\rate}{dE} dE \propto \volt^\alpha \; .
\label{rateV}
\ee
Thus, in the model without phonons, 
the tunneling exponent is $\alpha = 1/\nu$---the prediction 
of the chiral CFT.\cite{Wen}

\section{Correction to the tunneling exponent} \label{sect:corr}
We are interested in corrections to the results (\ref{rateE2}) and (\ref{rateV})
due to the piezoelectric coupling (\ref{Hp}). 
The full Green function $G$ of the chiral boson
is given by the following Fourier transform
\be
G(x, \tau) = \int \frac{d\Omega dk_x}{(2\pi)^2} e^{-i \Omega \tau + i k_x x} 
F(\Omega, k_x) \; ,
\label{full}
\ee
where
\be
F^{-1} = 2\kappa k_x (i\Omega + \cm k_x)
- h^2 k_x^2 \int \frac{d^2 k_\perp}{(2\pi)^2} \frac{1}{\Omega^2 + \vph^2 k^2} \; ,
\label{F}
\ee
$k_\perp = (k_y, k_z)$, and $k^2 = k_x^2 + k_\perp^2$.
Note that
it is the coefficient of $\ln\tau$ in the boson Green function (at $x=0$)
that determines the tunneling exponent $\alpha$. Consequently, we are interested 
in logarithmic contributions to $G$.

Formally, Eq. (\ref{full}) is the exact solution to our problem. However, it is
not easy to analyze. Alternatively, it can be used to obtain a perturbative 
expansion of $G$, i.e., an expansion in powers of $h^2$:
\be
G(x, \tau) = G_0(x, \tau) + G_1(x,\tau) + \ldots \; .
\ee
The leading term is the free propagator (\ref{G0}), and the next-to-leading term 
equals
\be
G_1(x, \tau) = - \frac{h^2}{4\kappa^2} \int \frac{d^3 k d\Omega}{(2\pi)^4} 
 \frac{e^{-i \Omega \tau + i k_x x}}{(\Omega - i c_- k_x)^2 (\vph^2 k^2 + \Omega^2)} 
\; .
\ee
Since the coupling $h^2$ is relatively small, we first concentrate on this term;
effects of the higher-order corrections are discussed in the next section.

Writing 
\be
G_1(x, \tau) = \frac{h^2}{16\pi^2 \kappa^2 \vph^2} I(x,\tau) \; ,
\label{G1I}
\ee
we find three types
of logarithmic terms in the integral $I$:  $I\approx  I_1 + I_2 + I_3$;
for $\tau > 0$
\ba
I_1 & = & - \frac{\tau}{\cm\tau + ix} \ln [\Lambda  (\cm \tau + i x ) ] \; , 
\nonumber \\
I_2 & = & \frac{\cm}{\vph^2 - \cm^2} \ln\frac{L}{\cm \tau + i x} \; , 
\nonumber \\
I_3 & = & - \half \sum_{s=\pm 1} \frac{1}{\vph + s \cm} \ln \frac{L}{\vph\tau - isx}
\; ; \nonumber
\ea
$\Lambda$ is the ultraviolet (momentum) cutoff.
The approximate sign in the expression for $I$ means that only terms that give
rise to $\ln\tau$ are kept.

Setting $x=0$, we obtain, to logarithmic accuracy
\be
G_1(x, \tau) = \frac{h^2 \nu^2}{\vph \cm (\vph + \cm)} \ln \tau \; .
\label{G1}
\ee
The corresponding correction to the tunneling exponent is 
\be
\Delta \alpha = - \frac{2 h^2}{\vph \cm (\vph + \cm)} \; ,
\label{Dalpha}
\ee
where the factor of 2 is due to the presence of two transverse 
polarizations of the phonon (the third,
longitudinal, polarization has larger velocity and gives a smaller contribution).
Note that the correction is always negative.

For estimates, we use the same expression for the piezoelectric coupling and the
same values of the parameters as in 
Ref. \onlinecite{Seelig&al}. In our present notation,
\be
h^2 = \frac{1}{\hbar \rho_M} \left( \frac{e e_{14}}{4\pi \epsilon} \right)^2
\label{h2}
\ee
where $\rho_M = 5.36~\mbox{g/cm$^3$}$, $e_{14} = 0.16~\mbox{C/m$^2$}$, and
$\epsilon = 13.2 \epsilon_0$; in this equation only, we have restored $\hbar$.
With these numbers, $h^2 = 5.39 \times 10^8$ m$^3$/s$^3$. Using also
$\vph = 3000$ m/s, we can rewrite Eq. (\ref{Dalpha}) as a function of a single
parameter, the ratio
\be
r = \cm / \vph \; .
\label{r}
\ee
We obtain
\be
\Delta \alpha = - \frac{0.04}{r (1 +r)} \; .
\label{Dalpha2}
\ee
Clearly, the smaller is $r$ the larger is the correction. While for $\cm=10^4$
m/s it is only 0.003 in the absolute value, for $\cm= 10^3$ m/s it is already
0.09. The latter number is roughly 
of the same order of magnitude as the discrepancy 
between the CFT prediction $\alpha = 3$ for $\nu =1 /3$
and the central values $\alpha = 2.65$--2.85
obtained experimentally\cite{Chang&al,Grayson&al} for different samples.
This number is also larger than the experimental uncertainty quoted for the 
experiment of Ref. \onlinecite{Chang&al}.

The most obvious way
to experimentally test the present theory is to reduce the plasmon speed via 
a capacitive coupling to external conductors. In that case, Eq. (\ref{Dalpha2}) 
predicts a further decrease in $\alpha$. This prediction, however, holds only 
insofar as corrections of higher orders in $h^2$ can be neglected, cf. the next
section.

\section{Higher-order corrections}
\label{sect:horder}
Expanding the exact Fourier transform (\ref{F}) in powers of $h^2$, we obtain
higher-order corrections to the plasmon Green function. In this section, we consider
the structure of the perturbation series for the case $\cm \ll \vph$, which is
the simplest. In this case, the main contribution to $G$ in a given order
can be found by dropping $\Omega^2$ in the integral over $k_\perp$. The $n$th-order
correction to the Green function at $\tau > 0$ becomes
\be
G_n(x,\tau) \approx \frac{\nu}{n!} \left(\frac{h^2 \nu \tau}{\vph^2}\right)^n
\!\! \int_0^\Lambda \!\!\!
dk_x  k_x^{n-1} e^{-(\cm \tau + ix) k_x} \ln^n\frac{\Lambda}{k_x} \; .
\label{Gn}
\ee
At $x = 0$ all these terms are positive, so their sum remains
the main contribution to the sum of the entire perturbation series.

From now on we concentrate on $x=0$. It turns out that 
the sum of (\ref{Gn}) over $n$ saturates at values of $n$ that are much smaller 
than $n_0 \equiv \Lambda \cm \tau$. For such $n$,  
the logarithm in (\ref{Gn}) can be considered a slowly varying function
of $k_x$, so that (for $n\neq 0$)
\be
G_n(x,\tau) \approx \frac{\nu}{n} 
\left(\frac{h^2 \nu}{\vph^2 \cm}\right)^n
\ln^n \frac{n_0}{n}  \equiv \frac{\nu}{n} \xi^n \ln^n \frac{n_0}{n} \; .
\label{Gn2}
\ee
The last equality defines the dimensionless coupling $\xi$.

It is somewhat more convenient to consider, instead of the sum of $G_n$, the sum 
of their derivatives
with respect to $n_0$. Thus, if we denote by $\Delta G$ the full correction to 
the Green function, $\Delta G = G - G_0$, then
\be
\frac{\partial \Delta G}{\partial n_0}  = \frac{\nu}{n_0}
\sum_{n=1}^{[n_0]}  \xi^n \ln^{n-1} \frac{n_0}{n} \; .
\label{deriv}
\ee
If $n$ is regarded as a continuous variable, the expression under the sum has a
maximum at $n=n_*$, where
\be
n_* = n_0 e^{-\frac{1+\xi}{\xi}} [ 1 + O(\frac{1}{n_*}) + O(\xi)] \; .
\label{n*}
\ee
We define a characteristic time $\tau_0$ by
\be
\tau_0^{-1} = \cm \Lambda \exp( -\frac{\vph^2 \cm}{h^2 \nu} - 1) \; .
\label{tau0}
\ee
Eq. (\ref{n*}) can now be written simply as $n_* \approx \tau/ \tau_0$. 
We see that $\tau_0$ is the infrared scale at which the perturbation theory 
breaks down.
At $\tau \ll \tau_0$, the sum over $n$ is essentially discrete and is well
approximated by the lowest-order term. On the other hand, if $\tau$ 
is significantly larger than $\tau_0$, many terms contribute. In that case,
$n$ is indeed quasi-continuous.

At even larger times, $\tau \gg \tau_0 / \xi$, the steepest descent condition 
becomes applicable near $n= n_*$, and using steepest descent we obtain
\be
\frac{\partial \Delta G}{\partial n_0}  = \frac{\nu}{n_0}
(2\pi n_* \xi)^{1/2} \exp [ \xi\tau/\tau_0 + O(n_* \xi^2) ] \; .
\label{Gsum}
\ee
The exponential growth of Eq. (\ref{Gsum}) at large $\tau$
indicates the presence of a plasmon state with energy
\be
E_p = - \frac{h^2 \nu}{\vph^2 \cm \tau_0} \; .
\label{Ep}
\ee
The same conclusion can be reached by looking directly at the resummed 
expression (\ref{F}) or, more precisely, its real-time version obtained by
replacing $\Omega$ with $-i\omega$. (In the presence of a state
with a negative energy, the rotation to Euclidean frequencies needs
to be redefined.) Neglecting again the frequency dependence of the integral 
over $k_\perp$, we see that the interaction shifts the pole of $F$ from
$\omega = -\cm k_x$ to $\omega = f(k_x)$, where 
\be
f(k_x) = -\cm k_x ( 1 - \xi \ln\frac{\Lambda}{|k_x|} ) \; .
\label{f}
\ee
This function has extrema at $k_x = \pm (\cm \tau_0)^{-1}$, where it equals
$\pm |E_p|$.

We interpret the presence of plasmon states with negative energies as a reflection
of the polaron effect---formation
of a bound state of an electron and the phonon field. This interpretation
is supported by the following estimate. Suppose we qualitatively 
describe the cumulative effect of the attractive interaction (\ref{U}) over time 
by the time-independent potential
\be
{\tilde U}(x) = \int U(x,\tau) d\tau = - \frac{\pi h^2}{\vph^2 |x|} \; .
\label{tU}
\ee
For an electron cloud with density distribution $n(x)$ and spatial size 
of order of the ``correlation length'' $\cm \tau_0$, the binding energy can
then be estimated as
\be
E_e = \half \int dx dy n(x) {\tilde U}(x-y) n(y) \sim 
- \frac{h^2\ln(\Lambda \cm \tau_0)}{\vph^2 \cm \tau_0}  \; .
\label{Ee}
\ee
An electron cloud can be thought of as containing of order 
$(1/\nu) \ln(\Lambda \cm \tau_0)$
plasmons (in the sense that this is how many plasmons are typically produced when an
electron-hole pair annihilates). Dividing Eq. (\ref{Ee}) by this number, 
we obtain an estimate of energy per plasmon in agreement with Eq. (\ref{Ep}).

A corollary to this argument is that any infrared effects associated with 
production of plasmons should saturate at time scales of order $\tau_0$
(equivalently, length scales of order $\cm \tau_0$, the size of the polaron).
We therefore expect that in the extreme infrared, at biasing energies 
\be
E \sim |E_e| \sim \frac{1}{\nu\tau_0} \; ,
\label{cross}
\ee
the system, at any $\nu$, will cross over to the normal Fermi-liquid 
behavior. 

We now estimate the timescale $\tau_0$ for $\nu=1/3$ and $\nu = 1$. 
The maximal phonon momentum $\Lambda$ is
$\Lambda = \omega_D / \vph$, where $\omega_D$
is the Debye frequency; for GaAs, $\omega_D = 345$ K.\cite{Adachi} 
Given the uncertainty introduced by 
the directional averaging of $h$ and by the imprecise knowledge of $r=\cm/\vph$, 
we prefer to extract the entire combination
$y= h^2/ \vph^2 \cm$, occurring in Eq. (\ref{tau0}),
directly from the experimentally measured values of the tunneling exponent.
In the limit $\cm \ll \vph$, the lowest order correction (\ref{Dalpha}) is simply
$\Delta \alpha = -2 y$.
Assuming (and confirming a posteriori) that the perturbation theory works well
for the $\nu=1/3$ state, we find that, e.g., the sample\cite{Chang&al} 
with $\alpha = 2.7$ corresponds to $y= 0.15$. For $\nu = 1/3$, 
Eq. (\ref{tau0}) then gives 
$\tau_0^{-1} \sim 10^{-9}r \omega_D$, which is unobservably small.
We conclude that for $\nu = 1/3$, in the experimentally relevant range of 
voltages and temperatures, 
the lowest-order result (\ref{Dalpha}) is reliable.

The situation is quite different for $\nu =1$. Using the same value of $y$,
we now obtain $\tau_0^{-1} = 14r$ $\mu$eV, close to the observable range 
of energies.
The lowest-order result (\ref{Dalpha}) predicts a negative correction
to the tunneling exponent at $\nu = 1$. 
Experimentally, no such negative correction has been
observed.\cite{Chang&al,Hilke&al} We see that this may be related to the breakdown 
of the perturbation theory for $\nu = 1$.

\section{Conclusion} \label{sect:concl}
We have computed the leading correction [Eq. (\ref{Dalpha})], 
due to the piezoelectric coupling to 3D phonons, to the current-voltage
exponent $\alpha$ for tunneling between a Fermi-liquid and 
a $\nu = 1/(2p + 1)$ QH edge.
The correction is always negative, in agreement with 
the experiments\cite{Chang&al,Grayson&al} on the $\nu= 1/3$ state, and
its magnitude depends on the value of the edge plasmon speed.

We have also shown that, in the experimentally relevant range of energies,
higher-order corrections for $\nu = 1/3$ are small and do not invalidate the
leading-order result, but for $\nu = 1$ they very well might. In neither case,
however, the leading-order result (\ref{Dalpha}) represents the true 
infrared behavior: at a sufficiently
low energy, higher-order effects will always become important. 

The author thanks A. Boyarsky, A. Chang, and V. Cheianov for discussions.

\end{document}